\documentclass[
reprint,
 amsmath,amssymb,
 aps,
prb,
floatfix,
]{revtex4-1}

\usepackage{graphicx}
\usepackage{dcolumn}
\usepackage{bm}
\usepackage{hyperref}
\usepackage[mathlines]{lineno}
\usepackage{commath}
\usepackage{physics}
\usepackage{color}
\usepackage{mathtools}

\begin{document}

\preprint{APS/123-QED}

\title{Reversible giant out-of-plane Rashba effect in two-dimensional Ga$XY$ ($X$= Se, Te; $Y$= Cl, Br, I) compounds for persistent spin helix}

\author{Siti Amalia Sasmito}%
\affiliation{Department of Physics, Universitas Gadjah Mada, Sekip Utara, BLS 21 Yogyakarta Indonesia.}%

\author{Muhammad Anshory}%
\affiliation{Department of Physics, Universitas Gadjah Mada, Sekip Utara, BLS 21 Yogyakarta Indonesia.}%

\author{Ibnu Jihad}%
\affiliation{Department of Physics, Universitas Gadjah Mada, Sekip Utara, BLS 21 Yogyakarta Indonesia.}%

\author{Moh. Adhib Ulil Absor}
\email{adib@ugm.ac.id} 
\affiliation{Department of Physics, Universitas Gadjah Mada, Sekip Utara, BLS 21 Yogyakarta Indonesia.}%

\date{\today}

\begin{abstract}
The coexistence of ferroelectricity and spin-orbit coupling (SOC) in noncentrosymmetric systems may allow for a nonvolatile control of spin degrees of freedom by switching the ferroelectric polarization through the well-known ferroelectric Rashba effect (FRE). Although the FER has been widely observed for bulk ferroelectric systems, its existence in two-dimensional (2D) ferroelectric systems is still very rarely discovered. Based on first-principles calculations, supplemented with $\vec{k}\cdot\vec{p}$ analysis, we report the emergence of the FRE in the Ga$XY$ ($X$= Se, Te; $Y$= Cl, Br, I) monolayer compounds, a new class of 2D materials having in-plane ferroelectricity. Due to the large in-plane ferroelectric polarization, a giant out-of-plane Rashba effect is observed in the topmost valence band, producing unidirectional out-of-plane spin textures in the momentum space. Importantly, such out-of-plane spin textures, which can host a long-lived helical spin mode known as a persistent spin helix, can be fully reversed by switching the direction of the in-plane ferroelectric polarization. Thus, our findings can open avenues for interplay between the unidirectional out-of-plane Rashba effect and the in-plane ferroelectricity in 2D materials, which is useful for efficient and non-volatile spintronic devices.
\end{abstract}

\pacs{Valid PACS appear here}
\maketitle

\section{INTRODUCTION}

During the last decade, spin-orbit coupling (SOC) has attracted increasing interest in various fields, including spintronics, quantum computing, topological matter, and cold atom systems\cite{Manchon, Varignon}. In particular, the SOC links the spin degree of freedom to the orbital motion of electrons in a solid without additional external magnetic field, thus playing an important role in semiconductor-based spintronics\cite{Manchon, Varignon, Ganichev}. For a system with a lack of inversion symmetry, the SOC induces an effective magnetic field, which results in spin-splitting bands and non-trivial spin textures in the momentum space, known as the Rashba \cite{Rashba} and Dresselhaus\cite{Dresselhaus} effects. The Rashba effect has been widely observed on a system having structural inversion asymmetry such as semiconductor quantum well \cite{Nitta, Caviglia}, surface heavy metal\cite{Koroteev, LaShell}, and several two-dimensional (2D) layered compounds\cite{Zhuang, Popovi, Absor_R, Affandi, Absor_Pol}, while the Dresselhaus effect occurs on a system hold bulk inversion asymmetries such as bulk zincblende\cite{Dresselhaus} and wurtzite semiconductors\cite{Wang_Dress}. 
 
Recently, ferroelectric materials have witnessed a surge of interest in the field of spintronics since they enable integration of the SOC and ferroelectricity through the well-known ferroelectric Rashba effect (FRE)\cite{Picozzi}. In such functionality, the spin textures of the spin-splitting bands can be fully reversed in a non-volatile way by switching the direction of the ferroelectric polarization. As such, the FER is very promising for spintronic devices implementing, for instant, tunneling anomalous and spin Hall effects \cite{Vedyayev, PMatos}. The FER was first predicted theoretically in bulk GeTe \cite{DiSante} and experimentally confirmed in GeTe thin-film\cite{Liebmann, Rinaldi}. After that, numerous candidates for FRE materials have been recently proposed, which mainly comes from the bulk metal-organic halide perovskite, including (FA)SnI$_{3}$ \cite{Stroppa, Kepenekian}, hexagonal semiconductors (LiZnSb)\cite{Narayan}, and oxides (KTaO$_{3}$\cite{Tao}, HfO$_{2}$\cite{TaoLL}, BiAlO$_{3}$\cite{daSilveira}). 

While the FRE has been widely studied for the bulk ferroelectric materials\cite{Picozzi, DiSante, Stroppa, Kepenekian, Narayan, Tao, TaoLL, daSilveira}, due to favorable spintronic applications in nanoscale devices \cite{Ahn2020, Han2016}, ultrathin two-dimensional (2D) materials supporting the FRE would be more desirable. However, the small thickness in the 2D materials may lose the FRE functionality since the ferroelectric polarization is suppressed by an enormous depolarizing field \cite{Junquera2003, Gao2017}. Recently, a new class of 2D materials exhibiting robust ferroelasticity and ferroelectricity has been reported, which comes from Ga$XY$ ($X$= Se, Te; $Y$= Cl, Br, I) monolayer (ML) compounds \cite{Zhou, ZhangShi}. These compounds are stable under room temperature exhibiting the large in-plane ferroelectricity \cite{Zhou, ZhangShi, Wu, Absor2021}. Moreover, due to the strong SOC in these materials, the large band splitting with tunable spin polarization in the conduction band minimum have recently been predicted \cite{Absor2021}. In addition to the observed large spin splitting in the Ga$XY$ ML compounds, the SOC induces doubly degenerate nodal loops featuring an hourglass type dispersion has also been reported \cite{Wu}. Considering the fact that the Ga$XY$ ML exhibits large in-plane ferroelectricity and strong SOC, it is expected that achieving the FRE in these materials is highly plausible, which is expected to be useful for spintronic applications. 

\begin{figure}
	\centering		
	\includegraphics[width=0.45\textwidth]{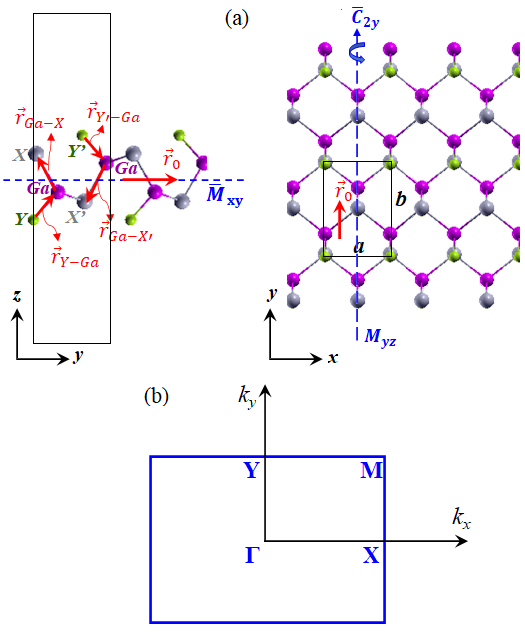}
	\caption{Atomic structure of the Ga$XY$ ML compounds corresponding to their symmetry operations viewed in the $x-y$ and $x-z$ planes, respectively, is presented. The unit cell of the crystal is indicated by the black lines and characterized by $a$ and $b$ lattice parameters in the $x$ and $y$ directions. The crystal is characterized by the glide reflection $\bar{M}_{xy}$ consisted of reflection about $z=0$ plane followed by $a/2$ translation along the $x$ axis and $b/2$ translation along the $y$ axis, the twofold screw rotation $\bar{C}_{2y}$ consisted of $\pi/2$ rotation around $y=b/2$ line followed by $a/2$ translation along the $x$ axis, and  the mirror reflection $M_{yz}$ around the $x=0$ plane. $\vec{r}_{Ga-X(X')}$ and $\vec{r}_{Y(Y')-Ga}$ are the vectors connected the Ga atom to chalcogen $X(X')$ atom and the halogen $Y(Y')$ atom to Ga atom, respectively in the unit cell. These vectors determine the distortion vector as indicated by $\vec{r}_{0}$. (b) first Brillouin zone of the Ga$XY$ ML compounds is shown, where high symmetry $\vec{k}$ points ($\Gamma$, $X$, $Y$, and $M$) are indicated.}
	\label{figure:Figure1}
\end{figure}

In this paper, through first-principles density-functional theory (DFT) calculations, complemented with $\vec{k}\cdot\vec{p}$ analysis, we predict the emergence of the FRE in the 2D Ga$XY$ ML compounds. We find that due to the large in-plane ferroelectric polarization in the Ga$XY$ ML, a giant out-of-plane Rashba effect is observed in the topmost valence band, exhibiting unidirectional out-of-plane spin textures in the momentum space. Importantly, such out-of-plane spin textures, which can host a long-lived helical spin mode known as a persistent spin helix (PSH), can be fully reversed by switching the direction of the in-plane ferroelectric polarization. Moreover, the physical mechanism of the FRE found in the present system is well analyzed within the framework of the $\vec{k}\cdot\vec{p}$ Hamiltonian model incorporating the in-plane ferroelectricity and point group symmetry of the crystal. Finally, a possible implication of the reversible spin textures of the present system for spintronics will be discussed.

\section{Computational details}

We have performed first-principles DFT calculations using the OpenMX code\cite{Ozaki, OpenmX, Ozakikino, Ozakikinoa}, based on norm-conserving pseudo-potentials and optimized pseudo-atomic localized basis functions. The exchange-correlation functional was treated within generalized gradient approximation by Perdew, Burke, and Ernzerhof (GGA-PBE)\cite{gga_pbe, Kohn}. The basis functions were expanded by linear combination of multiple pseudo atomic orbitals generated using a confinement scheme \cite{Ozaki,Ozakikino}, where two $s$-, two $p$-, two $d$-character numerical pseudo atomic orbitals were used. The accuracy of the basis functions as well as pseudo-potentials we used were carefully bench-marked by the delta gauge method \cite{Lejaeghere}.  

We applied a periodic slab to model the Ga$XY$ ML, where a sufficiently large vacuum layer (20 \AA) was applied in order to avoid the spurious interaction between slabs [Fig. 1(a)]. The $12\times10\times1$ $k$-point mesh was used to discretize the first Brillouin zone (FBZ) [Fig. 1(b)].  We adopted the modern theory of polarization based on the Berry phase (BP) method \cite{berry} implemented in the OpenMX code to calculate the ferroelectric polarization. During the structural relaxation, the energy convergence criterion was set to $10^{-9}$ eV. The lattice and positions of the atoms were optimized until the Hellmann-Feynman force components acting on each atom was less than 1 meV/\AA.  

The SOC was included self consistently in all calculations by using $j$-dependent pseudo potentials \citep{Theurich}. We calculated the spin textures by deducing the spin vector components ($S_{x}$, $S_{y}$, $S_{z}$) in the reciprocal lattice vector $\vec{k}$ from the spin density matrix\cite{Kotaka_2013}. The spin density matrix, $P_{\sigma \sigma^{'}}(\vec{k},\mu)$, were calculated using the following relation,  
\begin{equation}
\label{1}
P_{\sigma \sigma^{'}}(\vec{k},\mu)=\int \Psi^{\sigma}_{\mu}(\vec{r},\vec{k})\Psi^{\sigma^{'}}_{\mu}(\vec{r},\vec{k}) d\vec{r},
\end{equation}
where $\Psi^{\sigma}_{\mu}(\vec{r},\vec{k})$ is the spinor Bloch wave function. This methods has been successfully applied on our recent studies on various 2D materials\cite{Anshory2019, Absor2019, Absor2019b, Absor2020, Absor2021, Absor_JPCM}. 

\section{Results and Discussion}

\subsection{Atomic structure, symmetry, and ferroelectricity}

\begin{table*}
\caption{The optimized structural-related parameters and ferroelectric polarization are displayed. Here, the lattice parameters [$a$ (in \AA), $b$ (in \AA)], the bondlength between the Ga and chalcogen $X$($X'$) atoms [$|\vec{r}|_{Ga-X(X')}$ (in \AA)], the bondlength between the halogen $Y$($Y'$) and Ga atoms [$|\vec{r}|_{Y(Y')-Ga}$ (in \AA)], the magnitude of the distortion vector $|\vec{r}_{0}|$ (\AA), and  the in-plane electric polarization $P$ (in pC/m) obtained for each Ga$XY$ ML compounds are shown.} 
\begin{tabular}{c c c c c c  c } 
\hline\hline 
Ga$XY$ ML & $a$ (\AA)  & $b$ (\AA)   &  $|\vec{r}|_{Ga-X(X')}$ (\AA) &  $|\vec{r}|_{Y(Y')-Ga}$ (\AA)  & $|\vec{r}_{0}|$ (\AA)    & $P$ (pC/m) \\ 
\hline 
 GaSeCl &  3.87   &    5.53  & 2.47   &  2.23 &  0.23   & 478.9       \\ GaSeBr &  3.95   &    5.63  & 2.47   &  2.37 & 0.15   & 459.1       \\  
 GaSeI  &  4.17   &    5.93  & 2.49   &  2.60 & 0.08  & 352.5       \\   
 GaTeCl &  4.17   &    5.93  & 2.70   &  2.26 & 0.33  & 542.6       \\   GaTeBr &  4.26   &    6.08  & 2.71   &  2.37 & 0.28  & 530.1        \\  
 GaTeI  &  4.41   &    6.33  & 2.73   &  2.61 & 0.25   & 519.8       \\  
\hline\hline 
\end{tabular}
\label{table:Table 1} 
\end{table*}

First, we characterize the optimized structural parameters, symmetry of the crystal, and ferroelectricity of the Ga$XY$ ML compounds, where the atomic structure is displayed in Fig. 1(a). The crystal structure of the Ga$XY$ ML is noncentrosymmetric having a black-phosphorene-type structure belonging to $Pmn2_{1}$ space group \cite{Zhou, ZhangShi, Wu, Absor2021, Kniep}. For the convenience ofdiscussion, we choose the $x$ ($y$) axis to be along the zigzag (armchair) direction in the real space so that the reciprocal space is characterized by the FBZ as shown in Fig. 1(b). There are six atoms in the unit cell consisted of two Ga atoms, two chalcogen atoms (labeled by $X$ and $X'$), and two halogen atoms (labeled by $Y$ and $Y'$). These atoms are invariant under the following symmetry operations: (i) identity operation $E$, (ii) the glide reflection $\bar{M}_{xy}$ consisted of reflection about $z=0$ plane followed by $a/2$ translation along the $x$ axis and $b/2$ translation along the $y$ axis, where $a$ and $b$ is the lattice parameters of the crystal, (iii) the twofold screw rotation $\bar{C}_{2y}$ defined as $\pi/2$ rotation around $y=b/2$ line followed by $a/2$ translation along the $x$ axis, and (iv) the mirror reflection $M_{yz}$ around the $x=0$ plane. The optimized lattice parameters ($a$, $b$) for each Ga$XY$ ML compound are listed in Table 1. We find that due to the difference value between the $a$ and $b$ parameters, the crystal geometry of the Ga$XY$ ML is anisotropic, implying that these materials have different mechanical responses being subjected to uniaxial strain along the $x$- and $y$-direction similar to that observed on various group IV monochalcogenide\cite{Anshory2019, Kong2018, Liu2019}.   

The atomic structure of the Ga$XY$ ML can be viewed as Ga$X(X')$ ML surface functionalized by halogen $Y$ ($Y'$) atoms bonded to the Ga atoms forming a sandwiched structure with $Y$-Ga$X(X')$-$Y'$ sequence [see Fig. 1(a)]. We then introduce a distortion vector, $\vec{r}_{0}$, defined as 
\begin{equation}
\label{2}
\vec{r}_{0}=\vec{r}_{Ga-X} + \vec{r}_{Y-Ga} +\vec{r}_{Ga-X'}+\vec{r}_{Y'-Ga},
\end{equation}
where $\vec{r}_{Ga-X(X')}$ and $\vec{r}_{Y(Y')-Ga}$ are the vectors connected the Ga atom to chalcogen $X(X')$ atom and the halogen $Y(Y')$ atom to Ga atom, respectively, in the unit cell [see left side in Fig. 1(a)]. Here, the magnitude  $|\vec{r}|_{Ga-X(X')}$ and $|\vec{r}|_{Y(Y')-Ga}$ represent the Ga-$X$($X'$) and $Y$($Y'$)-Ga bond lengths, respectively. Due to the $M_{yz}$ mirror symmetry operation along the $y-z$ plane, we obtain that $\vec{r}_{0}\cdot\hat{x}=0$, while the screw operation $\bar{C}_{2y}$ implies that $\vec{r}_{0}\cdot\hat{z}=0$. Accordingly, $\vec{r}_{0}$ should be parallel to the in-plane $y$ direction and induces intrinsic spontaneous polarization along the $y$ direction. Generally, the optimized structures of the Ga$XY$ ML compounds show that the $|\vec{r}|_{Ga-X(X')}$ bond lengths are larger than the $|\vec{r}|_{Y(Y')-Ga}$ bond lengths [see Table I]. However, the $|\vec{r}|_{Y(Y')-Ga}$ bond lengths substantially increases for the compounds with the same chalcogen ($X$) atoms but have the heavier halogen ($Y$) atoms, thus decreasing the magnitude of the distortion vector, $|\vec{r}_{0}|$.  Therefore, the decreased in magnitude of the in-plane electric polarization is expected, which is in fact confirmed by our BP calculation results shown in Table I.  The existence of the in-plane ferroelectricity allows us to maintain the FRE in the Ga$XY$ ML compounds, which is expected to be observed due to the large SOC.

In the next section, we will show how the in-plane ferroelectricity plays an important role in the SOC and electronic properties of the Ga$XY$ ML compounds. 

\subsection{Spin-orbit coupled ferroelectric and spin textures}

We will start our analysis by deriving the general SOC Hamiltonian in the 2D systems having in-plane ferroelectricity. The SOC Hamiltonian is further analyzed for the Ga$XY$ ML compounds within the framework of the $\vec{k}\cdot\vec{p}$ Hamiltonian model using the method of invariant\cite{Winkler}. Finally, we discuss the important implication of the derived SOC Hamiltonian in terms of the spin splitting and spin textures involving to the in-plane ferroelectricity. 

The SOC occurs in solid-state materials when an electron 
moving at velocity $\vec{v}$ through an electric field $\vec{E}$ experiences an effective magnetic field due to the relativistic transformation of electromagnetic fields. A general form of the SOC Hamiltonian $H_{SO}$ can be expressed as:
\begin{equation}
H_{SO}=\vec{\Omega}\cdot\vec{\sigma},
\label{3}
\end{equation}
where $\vec{\sigma}=(\sigma_{x},\sigma_{y},\sigma_{z})$ are the Pauli matrices and $\vec{\Omega}$ is a wave-vector dependent spin-orbit field (SOF) that is simply written as
\begin{equation}
\vec{\Omega}(\vec{k})= \alpha \hat{E}\times\vec{k},
\label{4}
\end{equation}
where $\alpha$ is the strength of the SOC that is proportional to the magnitude of the electric field, $|\vec{E}|$, $\hat{E}$ denotes the electric filed direction, and  $\vec{k}$ is the wave vector representing the momentum electron. The $H_{SO}$ is invariant under time reversal symmetry operations, $T$, so that the following relation holds, $T H_{SO} T^{-1}=-\vec{\Omega}(-\vec{k})\cdot \vec{\sigma}=H_{SO}$. Accordingly, the SOF is a odd in wave vector $\vec{k}$, i.e. $\vec{\Omega}(-\vec{k})=-\vec{\Omega}(\vec{k})$, which also depends on the crystal symmetry of the system. 

Lets us consider the general 2D systems having in-plane ferroelectricity, where we assumed that the spontaneous in-plane electric polarization being oriented along the in-plane $y$-direction. In this case, an effective electric field is induced, which is also oriented along the $y$-directions, $\vec{E}=E \hat{x}$. Due to the 2D nature of the systems, we have $\vec{k}=k_{x}\hat{x}+k_{y}\hat{y}$ for the wave vector $\vec{k}$, and by using the explicit form of the effective electric field $\vec{E}$, we find that the SOF $\vec{\Omega}$ in Eq. (\ref{4}), can be expressed as  
\begin{equation}
\vec{\Omega}=\alpha k_{x} \hat{z}.
\label{5}
\end{equation}
We can see that for the 2D systems having in-plane ferroelectricity, the SOF is enforced to be unidirectional in the out-of-plane direction. Inserting the Eq. (\ref{4}) to the Eq. (\ref{3}), we find that
\begin{equation}
H_{SO}=\alpha k_{x}\sigma_{z}.
\label{6}
\end{equation}
The Eq. (\ref{6}) clearly shows that the $H_{SO}$ is characterized only by one component of the wave vector $k_{x}$  and the out-of-plane spin vector $\sigma_{z}$, yielding a unidirectional out-of-plane Rashba effect.

\begin{table}[ht!]
\caption{Transformation rules for the wave vector $\vec{k}$ and spin vector $\vec{\sigma}$ under the considered point-group symmetry operations. Time-reversal symmetry, implying a reversal of both spin and momentum, is defined as $T=i\sigma_{y}K$, where $K$ is the complex conjugation, while the point-group operations are defined as $\hat{C}_{2y}=i\sigma_{y}$, $\hat{M}_{yz}=i\sigma_{x}$, and $\hat{M}_{xy}=i\sigma_{z}$. The last column shows the invarian terms, where the underlined term are invariant under all symmetry operations.} 
\centering 
\begin{tabular}{c c c c} 
\hline\hline 
  Symmetry  & $(k_{x}, k_{y})$   & $(\sigma_{x}, \sigma_{y}, \sigma_{z})$ & Invariants\\
  Operations  &  &  & \\ 
\hline 
$\hat{T}=i\sigma_{y}K$   &  $(-k_{x}, -k_{y})$  &  $(-\sigma_{x}, -\sigma_{y}, -\sigma_{z})$  & $k_{i}\sigma_{j}$  \\  
   &    &  & ($i,j=x,y,z$)  \\  
$\hat{C}_{2y}=i\sigma_{y}$   &  $(-k_{x}, k_{y})$  &  $(-\sigma_{x}, \sigma_{y}, -\sigma_{z})$ & $k_{x}\sigma_{x}$, \underline{$k_{x}\sigma_{z}$}, $k_{y}\sigma_{y}$ \\ 
$\hat{M}_{yz}=i\sigma_{x}$   &  $(-k_{x}, k_{y})$  &  $(\sigma_{x}, -\sigma_{y}, -\sigma_{z})$ & $k_{x}\sigma_{y}$, \underline{$k_{x}\sigma_{z}$}, $k_{y}\sigma_{x}$   \\
$\hat{M}_{xy}=i\sigma_{z}$   &  $(k_{x}, k_{y})$  &  $(-\sigma_{x}, -\sigma_{y}, \sigma_{z})$ &  \underline{$k_{x}\sigma_{z}$}, $k_{y}\sigma_{z}$ \\
\hline\hline 
\end{tabular}
\label{table:Table 2} 
\end{table}

The $H_{SO}$ in Eq. (\ref{6}) is also obtained by considering the wave-vector symmetry group at the high symmetry points in the first-Brillouin zone. Here, we assumed that only linear terms with respect to the wave vector $\vec{k}$ contribute to the $H_{SO}$. For the case of the Ga$XY$ ML compounds, the wave-vector symmetry group of the $Pmn2_{1}$ space group at the high symmetry points such as $\Gamma$, $X$, and $Y$ points, belongs to $C_{2v}$ point group\cite{Absor2021}, which has two mirror reflections about the $x-y$ plane ($M_{xy}$) and the $y-z$ plane ($M_{yz}$) as well as twofold rotation $C_{2y}$ around the $y$-axis. The transformation rules for the wave vector $\vec{k}$ and spin vector $\vec{\sigma}$ under the considered point-group symmetry operations are listed in Table II. By applying the methods of invariant\cite{Winkler}, we list all the invariant term of the $H_{SO}$ in the form of product between the $\vec{k}$ and $\vec{\sigma}$ components (see the right column in Table II) and select those specific terms which are invariant under all symmetry operations as indicated by the underlined terms in the right column in Table II. We find that only $k_{x}\sigma_{z}$ term of the $H_{SO}$ is invariant under all symmetry operations of the $C_{2v}$, which is identical to the $H_{SO}$ shown in Eq. (\ref{6}). 

\begin{figure}
	\centering
		\includegraphics[width=0.5\textwidth]{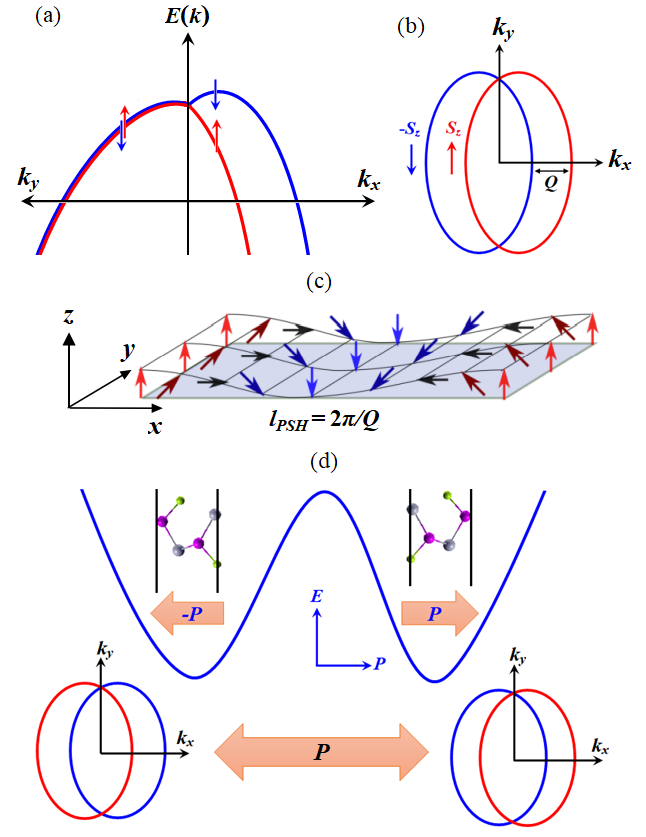}
	\caption{(a) Schematic view of band dispersion showing an anisotropic splitting around $\vec{k}=(0,0,0)$ and (b) the corresponding Fermi line are presented. The red and blue arrows indicate $S_{z}$ and $-S_{z}$ spin orientation in the momentum space, respectively. In Fig. (b), the Fermi line has the shifted two pair loops characterized by the shifting wave vector $\vec{Q}$ along the $k_{x}$ direction where the spins are persisting in the fully out-of-plane direction, resulting in the formation of the persistent spin textures (PST). (c) Schematic view of the persistent spin helix (PSH) mode emerge under the PST formation with the wavelength of $l_{PSH}=2\pi/|\vec{Q}|$. (d) Schematic correlation between the in-plane ferroelectricity and the spin textures is shown. The two stable ferroelectric phases of the 2D materials having opposite in-plane ferroelectric polarization are indicated. The switching spin textures are expected by reversing the in-plane ferroleectric polarization.}
	\label{fig:Figure2}
\end{figure}

Next, we characterize low energy properties of the present system involving the $H_{SO}$ term of Eq. (\ref{6}). The effective $\vec{k}\cdot \vec{p}$ Hamiltonian $H$ including the kinetic and $H_{SO}$ terms can be expressed as
\begin{equation}
H=\frac{\hbar^{2}k^{2}}{2m^{*}}+\alpha k_{x}\sigma_{z}.
\label{7}
\end{equation}
Solving eigenvalue problem involving the Hamiltonian of Eq. (\ref{7}) leads to the eigenstates 
\begin{equation}
\Psi_{\vec{k}\uparrow}=e^{i\vec{k}_{\uparrow}\cdot\vec{r}}\begin{pmatrix}
1\\
0
\end{pmatrix}
\label{8}
\end{equation}
and
\begin{equation}
\Psi_{\vec{k}\downarrow}=e^{i\vec{k}_{\downarrow}\cdot\vec{r}}\begin{pmatrix}
1\\
0
\end{pmatrix},
\label{9}
\end{equation}
corresponding to the energy dispersion,
\begin{equation}
E_{\uparrow\downarrow}=\frac{\hbar^{2}k^{2}}{2m^{*}}\pm \alpha k_{x}.
\label{10}
\end{equation}
This dispersion indicates that a strongly anisotropic spin splitting occurs around the $\vec{k}=(0,0,0)$ point, i.e., the energy bands are lifted along $k_{x}$ direction but are degenerated along the $k_{y}$ direction [Fig. 2(a)].  Importantly, this dispersion is characterized by the shifting property, $E_{\downarrow}(\vec{k})=E_{\uparrow}(\vec{k}+\vec{Q})$, where the  $\vec{Q}$ is the shifting wave vector given by
\begin{equation}
\vec{Q}=\frac{2m^{*}\alpha}{\hbar^{2}}[1,0,0].
\label{11}
\end{equation}
The Eqs. (\ref{10}) and (\ref{11}) implies that a constant-energy cut shows two Fermi loops whose centers are displaced from their original point by $\pm\vec{Q}$ as schematically shown in Fig. 2(b).

The spin texture, which is $\vec{k}$-dependent spin configuration, is determined from the expectation values of the spin operators, i.e., $\vec{S}=(\hbar/2)\langle \psi_{\vec{k}}|\vec{\sigma}|\psi_{\vec{k}}\rangle$, where $\psi_{\vec{k}}$ is the electron's eigenstates. By using $\psi_{\vec{k}}$  given in Eqs. (\ref{8}) and (\ref{9}), we find that \begin{equation}
\vec{S}_{\pm}=\pm \frac{\hbar}{2}[0,0,1].
\label{12}
\end{equation}
This shows that the spin configuration in the $k$-space is locked being oriented in the out-of-plane directions as schematically shown in Fig. 2(b). Such a typical spin configuration forms a persistent spin textures (PST) similar to that observed for [110] Dresselhauss model\cite{Bernevig}. Previously, it has been reported that the PST is known to host a long-lived helical spin mode known as a persistent spin helix (PSH) \cite{Bernevig, Altmann, Schliemann}, enabling long-range spin transport without dissipation\cite{Bernevig, Altmann, Schliemann, kohda2012, walser2012, koralek2009}, and hence very promising for an efficient spintronic devices. 

The PSH arises when the SOF is unidirectional, preserving a unidirectional spin configuration in the $k$-space. When an electron moving in the real space is accompanied by the spin precession around the SOF, a spatially periodic mode of the spin polarization is generated. According to Eq. (\ref{5}), the magnitude of the effective magnetic field can be expressed as $B=2|\vec{\Omega}|/\gamma\hbar$, where $\gamma$ is the gyromagnetic ratio. Therefore, the angular frequency of the precession motion, $\omega$, can be calculated using the relation, $\omega=-\gamma B= 2\alpha k_{x}/\hbar$. The spin precession angle, $\theta$, around the $y$ axis at time $t$, is obtained by   $\theta=\omega t=2\alpha k_{x} t/\hbar$. At the same time, the traveling distance of the electron is given by $l=vt=\hbar k_{x}t/m^{*}$, where $v$ is the electron velocity. By eliminating $t$, we find that $\theta=2\alpha m^{*} l/\hbar$. When $\theta=2\pi$, we then obtain the wavelength of the PSH, $l_{PSH}$\cite{Bernevig}, 
\begin{equation}
l_{PSH}=\pi \hbar^{2}/(m^{*}\alpha).
\label{13}
\end{equation}
Furthermore, in term of the shifting wave vector $\vec{Q}$ defined in Eq. (\ref{11}), we can write the $l_{PSH}$ as
\begin{equation}
l_{PSH}=2\pi/|\vec{Q}|.
\label{14}
\end{equation}
A schematic picture of the PSH mode enforced by the unidirectional SOF is displayed in Fig. 2(c), where a spatially periodic mode of the spin polarization with the wavelength $l_{PSH}$ is shown. Such spin-wave mode protects the spins of electrons from decoherence through suppressing the Dyakonov-Perel spin relaxation mechanism\cite{Dyakonov} and renders an extremely long spin lifetime\cite{Bernevig, Altmann, Schliemann, kohda2012, walser2012, koralek2009}.   

Finally, we study the correlation between spin textures and ferroelectricity. Here, an important property called reversible spin textures holds, i.e., the direction of the spin textures is locked and switchable by reversing the direction of the spontaneous electric polarization. Fig. 2(d) shows a schematic view of the spin textured ferroelectric in the Ga$XY$ ML  compounds showing fully reversible out-of-plane spin textures. It is shown that switching the direction of the in-plane ferroelectric polarization from $\vec{P}$ to $-\vec{P}$ leads to reversing the direction of the out-of-plane spin textures from $z$- to $-z$-direction.

From the symmetry point of view, switching the electric polarization direction $\vec{P}$ is equivalent to the space inversion symmetry operation $I$, which changes the wave vector from $\vec{k}$ to $-\vec{k}$, but preserves the spin vector $\vec{S}$\cite{Kim2014}. Now, suppose that $|\psi_{\vec{p}}(\vec{k})\rangle$ is the Bloch wave function of the crystal with electric polarization $\vec{P}$. The inversion symmetry operation $I$ on the Bloch wave function hold the following relation, $I|\psi_{\vec{P}}(\vec{k})\rangle=|\psi_{-\vec{P}}(-\vec{k})\rangle$. Applying the time-reversal symmetry $T$ brings $-\vec{k}$ back to $\vec{k}$ but flip the spin vector $\vec{S}$, thus $TI|\psi_{\vec{P}}(\vec{k})\rangle=|\psi_{-\vec{P}}(\vec{k})\rangle$. The expectation values of spin operator $\expval{S}$ can be further expressed in term of $\vec{P}$ and $\vec{k}$ vectors as
\begin{equation}
\label{15}
\begin{split}
\langle\vec{S}\rangle_{-\vec{P},\vec{k}} & =\langle\psi_{-\vec{P}}(\vec{k})|\vec{S}|\psi_{-\vec{P}}(\vec{k})\rangle\\
& = \langle\psi_{\vec{P}}(\vec{k})|I^{-1}T^{-1}\vec{S}TI|\psi_{\vec{P}}(\vec{k})\rangle\\
& = \langle\psi_{\vec{P}}(\vec{k})|(-\vec{S})|\psi_{\vec{P}}(\vec{k})\rangle\\
& = \langle-\vec{S}\rangle_{\vec{P},\vec{k}},
\end{split}
\end{equation}
which clearly shows that the spin directions is fully reversed by switching the direction of the electric polarization $\vec{P}$.

In the next section, we implement these general description of the spin-orbit coupled ferroelectric to discuss our results from the first-principles DFT calculations on various Ga$XY$ ML compounds.

\begin{figure*}
	\centering
		\includegraphics[width=1.0\textwidth]{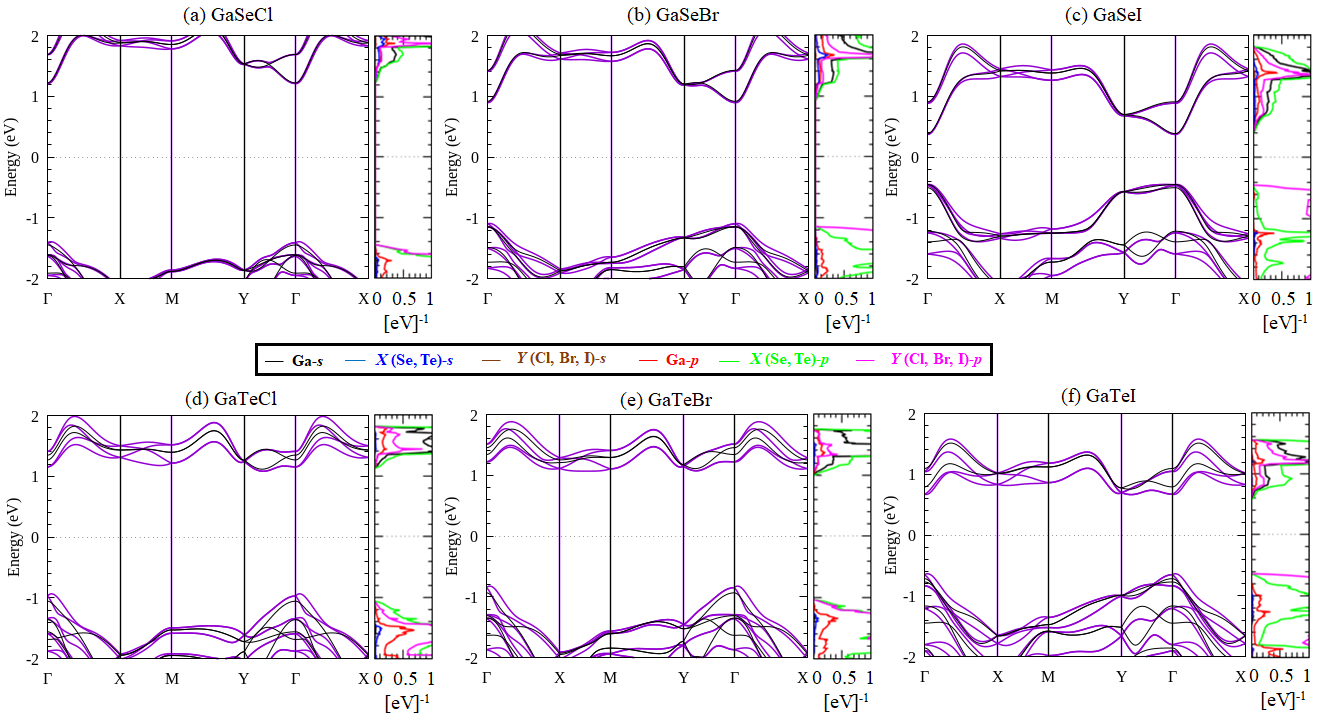}
	\caption{Electronic  band  structures  calculated  with  (purple  lines)  and  without  (black  lines)  including  the  SOC  for  various Ga$XY$ ML compounds: (a) GaSeCl, (b) GaSeBr, (c) GaSeI, (d) GaTeCl, (e) GaTeBr, and (d) GaTeI. Partial density of states (PDOS) projected to the atomic orbitals is also shown.  In the PDOS, the black, blue, yellow, red, green, and pink lines indicate the Ga-$s$, $X$ (Se, Te)-$s$, Ga-$p$, $X$ (Se, Te)-$p$, and $Y$ (Cl, Br, I)-$p$ orbitals, respectively.}
	\label{fig:Figure3}
\end{figure*}

\subsection{First-principles DFT analyses}

Figure 3 shows the electronic band structure of the Ga$XY$ ML compounds calculated along the selected $\vec{k}$ paths in the FBZ corresponding to the density of states (DOS) projected to the atomic orbitals. It is found that the Ga$XY$ ML compounds are semiconductors with direct or indirect band gaps depending on the chalcogen ($X$) atoms. In the case of the GaSe$Y$ MLs, the electronic band structure shows a direct bandgap where the valence band maximum (VBM) and conduction band minimum (CBM) is located at the $\Gamma$ point [Figs. 3(a)-3(c)]. The VBM at the $\Gamma$ point retains for the case of the GaTe$Y$ MLs but the CBM shifts to the $k$ point along the $\Gamma-Y$ line, resulting in an indirect bandgap [Figs. 3(d)-3(f)]. We find that the band gap significantly decreases for the compounds with the same chalcogen $X$ atoms but has the larger $Z$ number of the halogen ($Y$) atoms. For an instant, the calculated bandgap for the GaTeCl ML is 2.17 eV under GGA level, which is much larger than that for the GaTeI ML (1.10 eV). Our calculated DOS projected to the atomic orbitals confirmed that the VBM is mostly dominated by the contribution of the chalcogen $X$-$p$ orbital with a small admixture of the Ga-$p$ and halogen $Y$-$p$ orbitals, while the CBM is mainly originated from the Ga-$s$ orbital with a small contribution of Ga-$p$, chalcogen $X$-$p$ and halogen $Y$-$p$ orbitals [Figs. 3(a)-(f)].

\begin{figure*}
	\centering		
	\includegraphics[width=1.0 \textwidth]{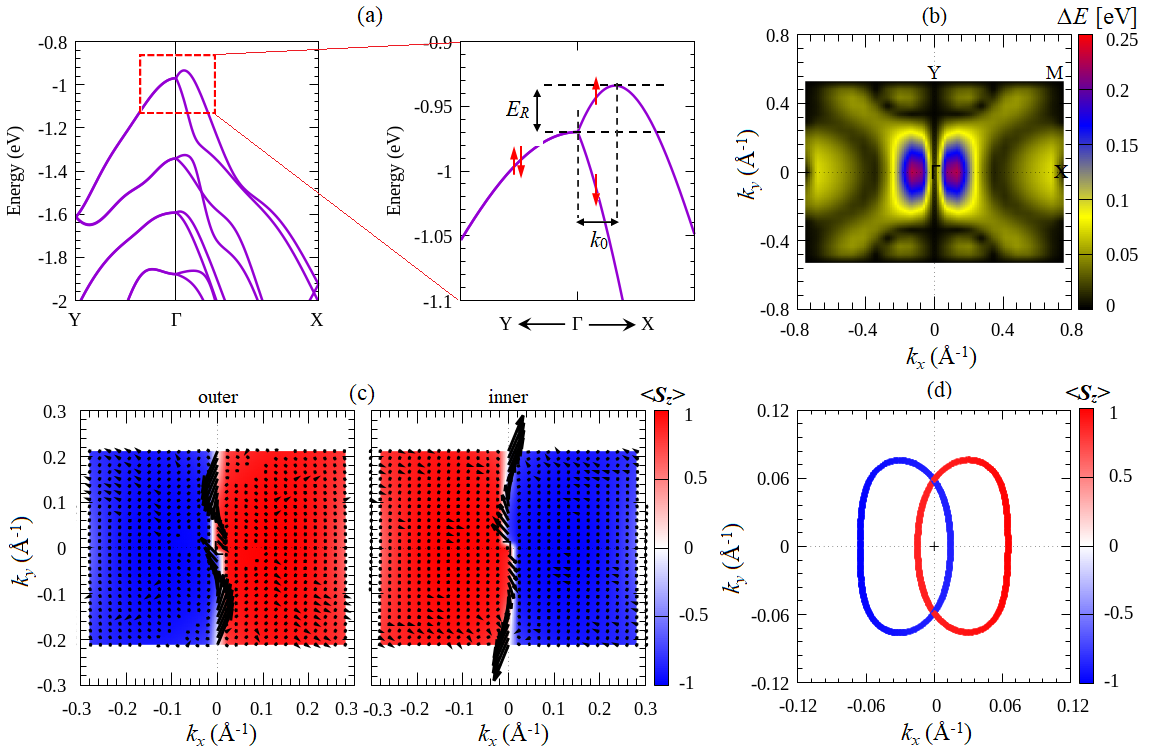}
	\caption{(a) Electronic  band  structures  calculated  with the  SOC around the VBM along $Y-\Gamma-X$ line for the GaTeCl ML as representative example of the Ga$XY$ ML compounds. The spin-split bands at the VBM around the $\Gamma$ point is highlighted. (b) Momentum-resolved map of the spin-splitting energy calculated along the entire of the first Brillouin zone is shown. The color bars in Fig. 4(b) shows the spin-splitting energy $\Delta E=|E(\vec{k}\uparrow)-E(\vec{k}\downarrow)|$, where $E(\vec{k}\uparrow)$ and $E(\vec{k}\downarrow)$ are the energy for the $\vec{k}$ bands with spin up and spin down, respectively. (c) Spin textures projected to the $k$-space for the upper and lower bands at the VBM around the $\Gamma$ point are shown. Here, color bars represent expectation values of the out-of-plane spin component $\left\langle S_{z}\right\rangle$. (d) $\left\langle S_{z}\right\rangle$ projected to the Fermi line calculated at constant energy cut of 1 meV below the degenerate state at the VBM around the the $\Gamma$ point.}
	\label{figure:Figure4}
\end{figure*}

Introducing the SOC, however, strongly modifies the electronic band structures of the Ga$XY$ ML compounds. Here, we observed a significant band splitting produced by the SOC due to the lack of the inversion symmetry, which is mainly visible at the $k$ bands along the $\Gamma-X-M$ symmetry lines [Figs. 3(a)-(f)]. However, along the $\Gamma-Y$ line in which the wave vector $\vec{k}$ is parallel to the effective electric field associated with the in-plane ferroelectric polarization, the bands are double degenerated. Since the electronic states near the Fermi level are important for transport carriers, we then focused our attention on the bands near the VBM. Fig. 4(a) shows the calculated band structure along the $Y-\Gamma-X$ line around the VBM for GaTeCl ML as a representative example of the Ga$XY$ ML compounds. At the $\Gamma$ point, the electronic states at the $\Gamma$ point are double degenerated due to time reversibility. This doublet splits into singlet when considering the bands $\vec{k}$ along the $\Gamma-X$ line. However, the doublet remains for the $\vec{k}$ along the $\Gamma-Y$ line, which is protected by the $\bar{C}_{2y}$ screw rotation and the $\bar{M}_{xy}$ glide mirror reflection. Accordingly, a strongly anisotropic splitting is clearly observed around the $\Gamma$ point as highlighted by the red line in Fig. 4(a), which is in good agreement with the energy dispersion shown in Eq. (\ref{10}) as well as Fig. 2(a). 

We noted here that the remaining band degeneracy at the wave vector $\vec{k}$ along the $\Gamma-Y$ line can be explained in term of the symmetry analysis. Since the wave vector $\vec{k}$ at the $\Gamma-Y$ line is invariant under $\bar{C}_{2y}$ and $\bar{M}_{xy}$ symmetry operations, the folowing relation holds, $\bar{M}_{xy}\bar{C}_{2y}=-e^{-ik_x}\bar{C}_{2y}\bar{M}_{xy}$, where the minus sign comes from the fact that both $\bar{C}_{2y}$ and $\bar{M}_{xy}$ operators are anti-commutative, $\left\{\bar{C}_{2y},\bar{M}_{xy}\right\}=0$ due to the anti-commutation between $\sigma_{y}$ and $\sigma_{z}$ spin rotation operators, $\left\{\sigma_{y},\sigma_{z}\right\}=0$. Supposed that $\left|\psi_{g}\right\rangle$ is an eigenvector of  $\bar{M}_{xy}$ operator with the eigenvalue of $g$, we obtain that $\bar{M}_{xy}(\bar{C}_{2y}\left|\psi_{g}\right\rangle)=-g(\bar{C}_{2y}\left|\psi_g\right\rangle)$. This evident shows that both $\left|\psi_{g}\right\rangle$ and $\bar{C}_{2y}\left|\psi_{g}\right\rangle)$ states are distinct states degenerated at the same energy, thus ensuring the double degeneracy of the states at the wave vector $\vec{k}$ along the $\Gamma-Y$ line.
To further clarify the observed anisotropic splitting around the $\Gamma$ point, we show in Fig. 4(b) momentum-resolved map of the spin-splitting energy calculated along the entire of the FBZ. Consistent with the band structures, we identify the non-zero spin-splitting energy except for the bands $\vec{k}$ along the $\Gamma-Y$ line. Here, the largest splitting is observed at the $\vec{k}$ closed to the $\Gamma$ point at along the $\Gamma-X$ line, where the splitting energy up to 0.25 eV is achieved. Such value is comparable with the splitting energy observed on various 2D transition metal dichalcogenides $MX_{2}$ ($M$= Mo, W; $X$ = S, Se, Te) MLs [0.15 eV - 0.46 eV] \cite{Zhu2011, Affandi, Absor_Pol, Yao2017, Absor2016}. The large splitting energy observed in the present system is certainly sufficient to ensure proper function of spintronic devices operating at room temperature\cite{Yaji2010}.

\begin{table*}
\caption{Several selected PST systems in 2D materials and parameters characterizing the strength of the SOC ($\alpha$, in eV\AA) and the wavelength of the PSH mode ($l_{PSH}$, in nm).} 
\begin{tabular}{c c c c} 
\hline\hline 
  2D materials & $\alpha$ (eV\AA)  & $l_{PSH}$ (nm)  & Reference \\ 
\hline 
\underline{Ga$XY$ compounds} &     &        &    \\
GaSeCl &  1.2   &    2.89    &   This work \\
GaSeBr&  0.85   &    4.09    &   This work \\
GaSeI &  0.53   &    6.57    &   This work \\
GaTeCl&  2.65   &    1.20    &   This work \\
GaTeBr &  2.40   &   1.45    &   This work \\
GaTeI&  1.90   &    1.83    &   This work \\
\underline{Group IV Monochalcogenide} &     &        &    \\
(Sn,Ge)$X$ ($X$= S, Se, Te)  & 0.07 - 1.67 & $8.9\times10^{2}$ - 1.82  & Ref.\cite{Absor2019}  \\
Ge$XY$ ($X, Y$= S, Se, Te) & 3.10 - 3.93& 6.53 - 8.52  & Ref.\cite{Absor_JPCM}  \\ 
Layeted SnTe & 1.28 - 2.85 & 8.80 - 18.3  & Ref.\cite{Lee2020}  \\ 
Strained SnSe & 0.76 - 1.15 &  & Ref.\cite{Anshory2019}\\
SnSe-$X$ ($X$= Cl, Br, I)  & 1.60 - 1.76 & 1.27 - 1.41 & Ref.\cite{Absor2019b}\\
\underline{Defective transition metal dichalcogenides} &     &        &    \\
line defect in PtSe$_{2}$ & 0.20 - 1.14 & 6.33 -  28.19 & Ref.\cite{Absor2020}\\
line defect in (Mo,W)(S,Se)$_{2}$ & 0.14 - 0.26 & 8.56 - 10.18 & Ref.\cite{Li2019}\\
\underline{Other 2D ML} &     &        &    \\
WO$_{2}$Cl$_{2}$ &  0.90     &      & Ref.\cite{Ai2019}\\

\hline\hline 
\end{tabular}
\label{table:Table 3} 
\end{table*}

The nature of the anisotropic splitting around the $\Gamma$ point at the VBM is further analyzed by identifying the spin textures of the spin-split bands. As shown in Fig. 4(c), it is found that a uniform pattern of the spin textures is observed around the $\Gamma$ point, which is mostly characterized by fully out-of-plane spin components $S_{z}$ rather than the in-plane spin components ($S_{x}, S_{y}$). These spin textures are switched from $S_{z}$ to $-S_{z}$ when crossing at $k_{x}$=0 along the $\Gamma-Y$ line. Although we identified large in-plane spin components ($S_{x}, S_{y}$) in the $\Gamma-Y$ line, the net in-plane spin polarization vanishes, which is due to the equal population of the opposite in-plane spin polarization between the outer and inner branches of the spin split bands [see black arrows in Fig. 4(c)].  Such a pattern of the spin textures, which is similar to that observed on several 2D ferroelectric materials such as WO$_{2}$Cl$_{2}$ \cite{Ai2019} and various group IV monochalcogenide MLs \cite{Absor2019, Lee2020, Anshory2019, Absor_JPCM}, is strongly different from the in-plane Rashba-like spin textures reported on the widely studied 2D materials\cite{Absor_R, Affandi, Absor_Pol, Yao2017}. Moreover, the fully out-of-plane spin texture becomes clearly visible when measured at the constant energy cut of 1 meV below the degenerated states at the VBM around the $\Gamma$ point [Fig. 4(d)].  Here, two circular loops of the Fermi lines with the opposite $S_{z}$ spin components are observed, which are shifted along the $\Gamma-X$ ($k_{x}$) direction. The observed spin textures, as well as Fermi lines, are all consistent well with our $\vec{k}\cdot\vec{p}$ Hamiltonian model presented in Eq. (\ref{7}) and the schematic pictures shown in Figs. 2(a)-(b). Remarkably, the observed unidirectional out-of-plane spin textures in the present system lead to the PST\cite{Bernevig, Schliemann}, which can host a long-lived helical spin-wave mode through the PSH mechanism \cite{Bernevig, Altmann, Schliemann, kohda2012, walser2012, koralek2009}. 

The observed spin splitting and spin textures can be quantified by the strength of the SOC, $\alpha$, which is obtained from the unidirectional out-of-plane Rashba model given by Eq. (\ref{7}). Here, we can rewrite the energy dispersion of Eq. (\ref{10}) in the following form:
\begin{equation}
E(k)=\frac{\hbar^{2}}{2m^{*}}\left(|k|\pm k_{0}\right)+E_{R},
\label{16}
\end{equation}
where $E_{R}$ and $k_{0}$ are the shifting energy and the wave vector evaluated from the spin-split bands along the $\Gamma-X$ ($k_{x}$) line as illustrated in Fig. 4(a). Accordingly, the following relation holds,
\begin{equation}
\alpha=\frac{2E_{R}}{k_{0}}.
\label{17}
\end{equation}
Both $E_{r}$ and $k_{0}$ are important parameters to stabilize spin precession and achieve a phase offset for different spin channels in the spin-field effect transistor device\cite{Datta}. In table III, we summarize the calculated result of the SOC strength $\alpha$ in Table III, and compare this result with a few selected PST systems previously reported on several 2D materials. It is found that the calculated value of $\alpha$ for the GaTeCl ML is 2.65 eV\AA, which is the largest among the Ga$XY$ ML compounds. This value is comparable with that observed on the PST systems reported for several 2D group IV monochalcogenide including Ge$XY$ ($X,Y$ = S, Se, Te) MLs (3.10 - 3.93 eV\AA) \cite{Absor_JPCM}, layered SnTe (1.28 - 2.85 eV\AA) \cite{Lee2020}. However, the calculated value of $\alpha$ is much larger than that observed on the PST systems found in other class of 2D materials such as WO$_{2}$Cl$_{2}$ ML (0.90 eV\AA) \cite{Ai2019} and transition metal dichalcogenide $MX_{2}$ MLs with line defect such as PtSe$_{2}$ (0.20 - 1.14 eV\AA) \cite{Absor2020} and (Mo,W)$X_{2}$ ($X$=S, Se) (0.14 - 0.26 eV\AA) \cite{Li2019}. 

\begin{figure*}
	\centering		
	\includegraphics[width=1.0\textwidth]{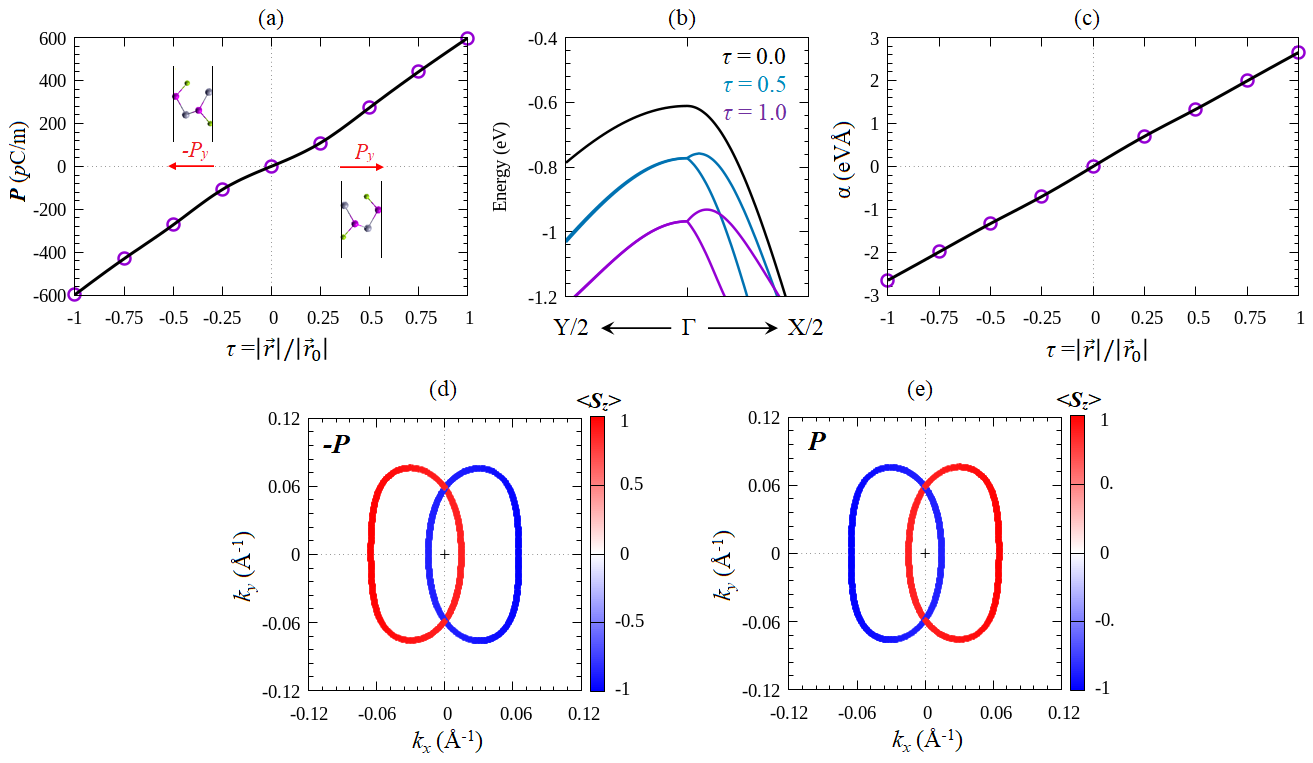}
	\caption{ Relation between polarization, spin splitting, and spin textures. (a) In-plane electric polarization $\vec{P}$ of the GaTeCl ML as a function of ferroelectric distortion $\tau$ is shown. The insert shows the optimized structure of the GaTeCl ML in the ferroelectric phase with $\vec{P}$ and $-\vec{P}$ polarization. $\tau$ is defined as the magnitude of the distortion vector $|\vec{r}|$ of the systems given in Eq. (\ref{1}) normalized by the magnitude of the distortion vector of the optimized ferroelectric phase, $|\vec{r}_{0}|$. Here $\tau=0$ represents the paraelectric phase and $\tau=1$ shows the optimized ferroelectric phase. (b) Band structure of the GaTeCl ML calculated along $Y-\Gamma-X$ line around the VBM as a function of the ferroelectric distortion $\tau$ is presented.  (c) The SOC strength $\alpha$ of the GaTeCl ML as a function of the ferroelectric distortion $\tau$ is presented.  Reversible out-of-plane spin orientation in GaTeCl ML calculated at constant energy cut of 1 meV below the degenerate state at the VBM around the $\Gamma$ point for the optimized ferroelectric phase with opposite in-plane electric polarization: (d) $-\vec{P}$ and (e) $\vec{P}$. }
	\label{figure:Figure5}
\end{figure*}

The emergence of the PST with large SOC strength $\alpha$ predicted in the present system indicates that the formation of the PSH mode with a substantially small wavelength $l_{PSH}$  of the spin polarization is achieved. Here, the wavelength $l_{PSH}$ can be estimated by using Eq. (\ref{13}) evaluated from the band dispersion along the $\Gamma-X$ line in the VBM [see the insert of Fig. 4(a)]. The resulting wavelength $l_{PSH}$ for all members of Ga$XY$ ML compounds are shown in Table III. In particular, we find a very small wavelength $l_{PSH}$ of the PSH mode for the GaTeCl ML (1.20 nm), which is the smallest over of all known 2D materials so far [see Table III]. Importantly, the small wavelength of the PSH mode observed in the present system is typically on the scale of the lithographic dimension used in the recent semiconductor industry\cite{Fiori2014}, which is possible to access the features down to the nanometers scale with sub-nanosecond time resolution by using near-field scanning Kerr microscopy. Thus, we concluded that that the present system is promising for miniaturization spintronics devices. 

Before summarizing, we highlighted the interplay between the in-plane ferroelectricity, spin splitting, and the spin textures in the Ga$XY$ ML compounds. Fig. 5(a) displayed the in-plane electric polarization as a function of the ferroelectric distortion, $\tau$. Here, $\tau$ is defined as the magnitude of the distortion vector $|\vec{r}|$ of the systems defined by Eq. (\ref{1}), which is normalized by the magnitude of the distortion vector of the optimized ferroelectric phase, $|\vec{r}_{0}|$. Therefore, $\tau=0$ represents the paraelectric phase, while $\tau=1$ shows the optimized ferroelectric phase as shown by the insert of Fig. 5(a). We can see that it is possible to manipulate the in-plane electric polarization $\vec{P}$ by distorting the atomic position [see Fig. 1(a)]. The dependence of the in-plane polarization on the ferroelectric distortion $\tau$ sensitively affects the spin-split bands at the VBM around the $\Gamma$ point as shown in Fig. 5(b). It is found that the splitting energy and the position of the VBM around the $\Gamma$ point strongly depend on the ferroelectric distortion, i.e., a decrease in $\tau$ substantially reduces the spin splitting energy while the position of the VBVM shifts up to be higher in energy around the $\Gamma$ point. Accordingly, the significant change of the SOC strength $\alpha$ is achieved, in which a linear trend of $\alpha$ as a function of $\tau$ is observed as shown in Fig. 5(c). Importantly, our results also show that the SOC strength $\alpha$ changes sign when the direction of the in-plane ferroelectric polarization $\vec{P}$ is switched, resulting in a full reversal of the out-of-plane spin textures shown in Figs. 5(d)-(e). Such reversible spin textures are agreed well with our symmetry analysis given by Eq. (\ref{15}), putting forward Ga$XY$ ML compounds as a candidate of the FER class of 2D materials exhibiting the PST, which is useful for efficient and non-volatile spintronic devices.  

\section{CONCLUSION}

In summary, we have investigated the emergence of the FRE in Ga$XY$ ($X$= Se, Te; $Y$= Cl, Br, I) ML compounds, a new class of 2D materials having in-plane ferroelectricity, by performing first-principles density-functional theory calculations supplemented with $\vec{k}\cdot\vec{p}$ analysis. We found that due to the large in-plane ferroelectric polarization, a giant unidirectional out-of-plane Rashba effect is observed in the spin-split bands around the VBM, exhibiting the unidirectional out-of-plane spin polarization persisting in the entirely FBZ. These persistent spin textures can host a long-lived persistent spin helix mode\cite{Bernevig, Altmann, Schliemann}, characterized by the large SOC strength and a substantially small wavelength of the helical spin polarization. Importantly, we observed fully reversible spin textures, which are achieved by switching the direction of the in-plane ferroelectric polarization, thus offering a possible application of the present system for efficient and non-volatile spintronic devices operating at room temperature. 

The reversible unidirectional out-of-plane Rashba effect found in the present study is solely enforced by the in-plane ferroelectricity and the non-symmorphic $Pnm2_{1}$ space group symmetry of the crystal. Therefore, it is expected that this effect can also be achieved on other 2D materials having similar crystal symmetry. Recently, there are numerous 2D materials that are predicted to have $Pnm2_{1}$ space group symmetry such as the 2D elemental group V (As, Sb, and Bi) MLs\cite{Pan2020, Xiao2018}. Due to the stronger SOC in these materials, the better resolution of the unidirectional out-of-plane Rashba effect is expected to be observed. Therefore, our prediction is expected to trigger further theoretical and experimental studies in order to find novel 2D ferroelectric systems supporting the unidirectional out-of-plane Rashba effect, which is useful for future spintronic applications.

\begin{acknowledgments}

This research was partly supported by RTA program (2021) supported by Universitas Gadjah Mada. Part of this research was supported by PDUPT (No.1684/UN1/DITLIT/DIT-LIT/PT/2021) and PD (No.2186/UN1/DITLIT/DIT-LIT/PT/2021) Research Grants funded by RISTEK-BRIN, Republic of Indonesia. The computation in this research was performed using the computer facilities at Universitas Gadjah Mada, Republic of Indonesia. 

\end{acknowledgments}

\bibliography{Reference1}


\end{document}